\documentstyle[11pt,fleqn,epsf]{article}
\newcommand\ion[2]{\hbox{#1\,{\sc #2}}}

\newcommand\teff{$ {\rm T_{eff}}$}
\newcommand\logt{$\log {\rm T_{eff}}$}
\newcommand\logg{$\log {\rm g}$}

\newcommand{\Msolar}{\mbox{\,$\rm M_{\odot}$}} 
\newcommand{\magpt}[2]{\mbox{$\rm #1\hspace{-0.25em}\stackrel{m}{.}
 \hspace{-1.0mm}#2$}}  

\newcommand{\litanf}{\begin{list}{}{\leftmargin=1.5cm \rightmargin=0cm 
\itemindent=-1.5cm \parsep=0cm \itemsep=0cm }}
\newcommand{\litend}{\end{list}}

\begin{document}

\title{Hot Stars in Globular Clusters\thanks{Based on observations
obtained at the ESO La Silla Observatory, the German-Spanish Calar Alto
Observatory and with the Hubble Space Telescope}}
\author{Sabine Moehler\\
\footnotesize Dr.\,Remeis-Sternwarte, 
\footnotesize Astronomisches Institut der Universit\"at Erlangen-N\"urnberg\\
{\footnotesize Sternwartstr.~7, 96049 Bamberg, Germany} \vspace{-5.5cm}\\
to appear in {\bf Reviews in Modern Astronomy Vol.~12}\\
	Astronomische Gesellschaft\\
 \vspace{4.0cm}\\
}
\date{}
\maketitle

\begin{abstract}

Blue horizontal branch and UV bright stars in several globular clusters are
analysed spectroscopically and the results are compared with predictions of
stellar evolutionary theory. We find that the distribution of temperatures and
surface gravities of the blue HB stars may be explained by the effects of
deep mixing. The masses derived for these stars are too low unless one uses
the long distance scale for globular clusters. First results on blue HB stars
in metal rich clusters are presented.

Analyses of hot UV bright stars in globular clusters uncovered a lack of 
genuine post-asymptotic giant branch stars which may explain the lack of
planetary nebulae in globular clusters seen by Jacoby et al. (1997). Abundance
analyses of post-AGB stars in two globular clusters suggest that gas and dust
may separate during the AGB phase.

\vspace*{1ex}
{\it ``As the series on metal-poor stars was originally conceived, this 
paper was to present the final solution to the appearance of the horizontal 
branch in the H-R diagram. Since that time, however, there have been 
several developments which obfuscate our understanding of these stars''
(Rood 1973)}
\end{abstract}

\section{Historical Background}

Today we know that galactic globular clusters are old stellar systems and 
people are therefore often surprised by the presence of hot stars in these 
clusters. As the following paragraphs will show hot stars have been known to
exist in globular clusters for quite some time:

Barnard (1900) reports the detection of stars in globular clusters that were
much brighter on photographic plates than they appeared visually: {\it ``Of
course the simple explanation of this peculiarity is that these stars, so
bright photographically and so faint visually, are shining with a much bluer
light than the stars which make up the main body of the clusters''}.

In 1915 Harlow Shapley started a project to obtain colours and magnitudes of 
individual stars in globular and open clusters (Shapley 1915a) hoping that 
{\it ``considerable advance can be made in our understanding of the internal
arrangement and physical characteristics''} of these clusters. In the first
globular cluster studied (M~3, Shapley 1915b) he found a double peaked
distribution of colours, with a red maximum and a blue secondary peak. He
noticed that - in contrast to what was known for field dwarf stars - the stars
in M~3 became bluer as they became fainter. Ten Bruggencate (1927, p.130) used
Shapley's data on M~3 and other clusters to plot magnitude versus colour 
(replacing luminosity and spectral type in the Hertzsprung-Russell diagram)
and thus produced the first colour-magnitude diagrams\footnote{Shapley (1930,
p.26, footnote) disliked the idea of plotting individual data points - he
thought that the small number of measurements might lead to spurious results.}
({\sc ``Farbenhelligkeitsdiagramme''}). From these colour-magnitude diagrams
(CMD's) ten Bruggencate noted the presence of a giant branch that became bluer
towards fainter magnitudes, in agreement with Shapley (1915b). In addition,
however, he saw a horizontal branch ({\sc ``horizontaler Ast''}) that parted
from the red giant branch and extended far to the blue at constant brightness.

\begin{figure}[ht]
\vspace*{11.cm}
\includegraphics{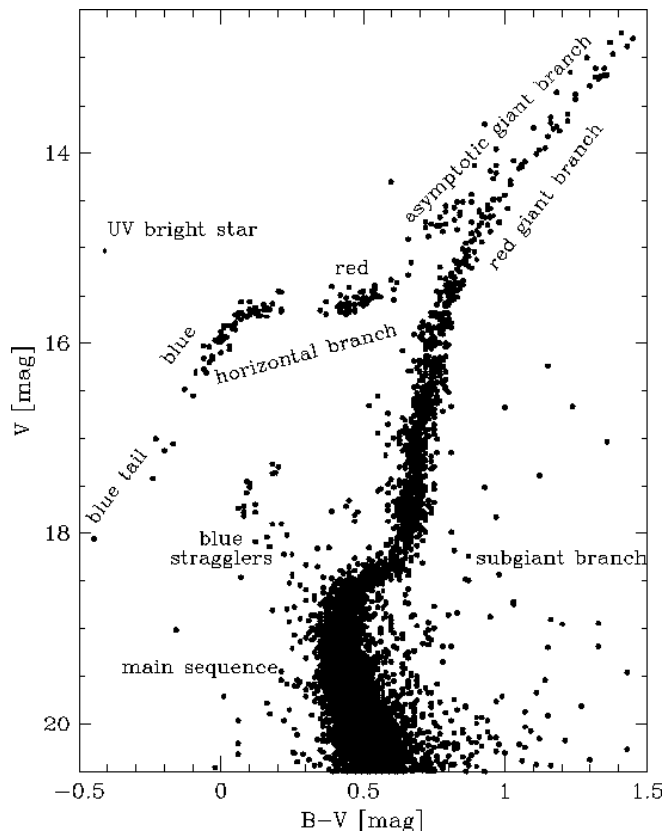}
\caption{Colour-magnitude diagram of M~3 (Buonanno et al. 1994) with 
the names of the principal sequences (some of which will be used in 
this paper).}
\label{cmd}
\end{figure}

Greenstein (1939) produced a colour-magnitude diagram for M~4 (again noting 
the presence of a sequence of blue stars at constant brightness) and mentioned
that {\it ``the general appearance of the colour-magnitude diagram of M4 is
almost completely different from that of any galactic {\rm (i.e. open)} 
cluster''}. He also noticed that - while main-sequence B and A type stars were
completely missing - there existed a group of bright stars above the
horizontal branch and on the blue side of the giant branch. Similar stars
appeared also in the CMD's presented by Arp (1955). As more CMD's of globular
clusters were obtained it became apparent that the horizontal branch
morphology varies quite considerably between individual clusters. The clusters
observed by Arp (1955) exhibited extensions of the blue horizontal branch
towards bluer colours and fainter visual magnitudes, i.e. towards hotter
stars\footnote{The change in slope of the horizontal branch is caused by the
decreasing sensitivity of B$-$V to temperature on one hand and by the
increasing bolometric correction for hotter stars on the other hand.} (see
 Fig.~\ref{cmd}). In some of Arp's CMD's (e.g. M~15, M~2) these {\bf blue
tails} are separated from the horizontal part by gaps (see also
 Fig.~\ref{cmds_obs}).

About 25 years after their discovery first ideas about the nature of the
horizontal branch stars began to emerge: Hoyle \& Schwarzschild (1955) were
the first to identify the horizontal branch with post-red giant branch (RGB)
stars that burn helium in their cores.

Sandage \& Wallerstein (1960) noted a correlation between the metal abundance
and the horizontal branch morphology seen in globular cluster CMD's: the
horizontal branch became bluer with decreasing metallicity. Faulkner (1966)
managed for the first time to compute zero age horizontal branch (HB) models
that qualitatively reproduced this trend of HB morphology with metallicity
(i.e. for a constant total mass stars become bluer with decreasing
metallicity) without taking any mass loss into account but assuming a rather
high helium abundance of Y = 0.35. Iben \& Rood (1970), however, found that
{\it ``In fact for the values of Y and Z most favored (Y $\ge$ 0.25
$\rightarrow$ 0.28, Z = $10^{-3} \rightarrow 10^{-4}$), individual tracks are
the stubbiest. We can account for the observed spread in color along the
horizontal branch by accepting that there is also a spread in stellar mass
along this branch, bluer stars being less massive (on the average) and less
luminous than redder stars. It is somewhat sobering to realize that this
conclusion comes near the end of an investigation that has for several years
relied heavily on aesthetic arguments against mass loss and has been guided by
the expectation of obtaining, as a final result, individual tracks whose color 
amplitudes equal the entire spread in color along the observed horizontal 
branches''}. In the same paper they found that {\it ``During most of the
double-shell-source phase, models evolve upwards and to the red along a
secondary giant branch\footnote{This secondary giant branch is called
asymptotic giant branch (AGB) later in the text and consists of stars with a
hydrogen and a helium burning shell.} that, for the models shown, approaches
the giant branch defined by models burning hydrogen in a shell.''} 

Comparing HB models to observed globular cluster CMD's Rood (1973) found that
an HB that {\it ``\ldots is made up of stars with the same core mass and
slightly varying total mass, produces theoretical c-m diagrams very similar to
those observed. \ldots A mass loss of perhaps 0.2~M$_\odot$ with a random
dispersion of several hundredths of a solar mass is required somewhere along
the giant branch.''} The assumption of mass loss also diminished the need for
very high helium abundances.

Sweigart \& Gross (1974, 1976) computed HB tracks including semi-convection 
and found that this inclusion considerably extends the temperature range
covered by the tracks. However, Sweigart (1987) noted that {\it ``For more 
typical globular cluster compositions, however, the track lengths are clearly
too short to explain the observed effective temperature distributions along
many HB's, and thus semiconvection does not alleviate the need for a spread in
mass (or some other parameter), a point first emphasized by Rood (1973)''}.

Caloi (1972) investigated zero age HB locations of stars with very low
envelope masses ($\le$ 0.02~\Msolar; extended or {\bf extreme HB} = EHB) and
found that they can be identified with subdwarf B stars in the field
(Greenstein 1971). Sweigart et al. (1974) and Gingold (1976) studied the
post-HB evolution and found that -- in contrast to the more massive blue HB
stars -- EHB models do not ascend the second (asymptotic) giant branch (AGB). 

Thus our current understanding sees {\bf blue horizontal branch stars} as stars 
that burn helium in a core of about 0.5~\Msolar\ and hydrogen in a shell. Their
hydrogen envelopes vary between $\ge$ 0.02~\Msolar\ (less massive envelopes
belong to EHB stars which do not have any hydrogen shell burning) and 0.1 --
0.15~\Msolar. Depending on the mass of their hydrogen envelopes they evolve to
the asymptotic giant branch (BHB stars) or directly to the white dwarf domain
(EHB stars, AGB manqu\'e stars). For a review see Sweigart (1994).

But blue horizontal branch stars are neither the brightest nor the bluest stars 
in globular clusters: Already Shapley (1930, p.30) remarked that {\it
``Occasionally, there are abnormally bright blue stars, as in Messier~13, but
even these are faint absolutely, compared with some of the galactic B
stars''}. This statement refers to stars like those mentioned by Barnard
(1900) which in colour-magnitude diagrams lie above the horizontal branch and
blueward of the red giant branch (see Fig.~\ref{cmd}). This is also the region
where one would expect to find central stars of planetary nebulae, which are,
however, rare in globular clusters: Until recently Ps1 (Pease 1928), the
planetary nebula in M~15 with its central star K~648, remained the only such
object known in globular clusters (see also Jacoby et al. 1997). 

The bright blue stars are clearly visible in the colour-magnitude diagrams of 
Arp (1955). Apart from analyses of individual stars like vZ~1128 in M~3 (Strom
\& Strom 1970, and references therein) and Barnard~29 in M~13 (Traving 1962,
Stoeckley \& Greenstein 1968) the first systematic work was done by Strom et
al. (1970). All stars analysed there show close to solar helium content,
contrary to the blue horizontal branch stars, which in general are depleted in
helium (Heber 1987). Strom et al. identified the brightest and bluest stars
with models of post-AGB stars (confirming the ideas of Schwarzschild \& H\"arm 
1970) and the remaining ones with stars evolving from the horizontal branch
towards the AGB. This means that all of the stars in this study are in the
double-shell burning stage. Zinn et al. (1972) performed a systematic search
for such stars using the fact that they are brighter in the U band than all
other cluster stars. This also resulted in the name {\bf UV Bright Stars} for
stars brighter than the horizontal branch and bluer than the red giant branch.
Zinn (1974) found from a spectroscopic analysis of UV bright stars in 8 
globular clusters {\it ``a strong correlation between the presence of supra-HB 
stars in a globular cluster and the presence of HB stars hotter than 
log~${T_{eff}}$ = 4.1''}. Harris et al. (1983) extended the compilation of UV
bright stars in globular clusters and de Boer (1987) gave another list of UV
bright stars in globular clusters, together with estimates of effective
temperatures and luminosities.

De Boer (1985) found from analyses of IUE spectra of UV bright stars in 7 
globular clusters that their contribution to the total cluster intensity ranges
{\it ``from, on average, over 50\% at 1200~\AA\ to a few percent at 
3000~\AA.''} Most of the UV bright stars found in ground based searches are
cooler than 30,000~K, although theory predicts stars with temperatures up to
100,000~K (e.g. Sch\"onberner 1983) The ground based searches, however, are
biased towards these cooler stars due to the large bolometric corrections for
hotter stars. It is therefore not very surprising that space based searches in
the UV (Ultraviolet Imaging Telescope, Stecher et al. 1997) discovered a 
considerable number of additional {\em hot} UV bright stars in a number of 
globular clusters.

Space based observatories also contributed a lot of other information about
hot stars in globular clusters: UIT observations showed the unexpected presence
of blue HB stars in metal-rich globular clusters like NGC~362 (Dorman et al.
1997) and 47~Tuc (O'Connell et al. 1997). At about the same time Hubble Space
Telescope (HST) observations of the core regions of globular clusters showed
long blue tails in metal-rich bulge globular clusters (Rich et al. 1997).
Observations of the very dense core regions of globular clusters show that the
colour-magnitude diagrams seen there may differ considerably from those seen
in the outer regions of the same clusters (e.g. Sosin et al. 1997). The most
recent addition to the family of hot stars in globular clusters are the white 
dwarfs seen in HST observations of M~4 (Richer et al. 1995, 1997), NGC~6752
(Renzini et al. 1996) and NGC~6397 (Cool et al. 1996), which unfortunately are
at the very limit for any spectroscopic observations even with 10m class
telescopes.

The interest in old hot stars like blue horizontal branch and UV bright stars
has been revived and extended by the discovery of the UV excess in elliptical
galaxies (Code \& Welch 1979; de Boer 1982) for which they are the most likely
sources (Greggio \& Renzini 1990, Brown et al. 1997). 

\section{Spectroscopic Analysis Methods} Much of the discussion and findings
described above are based solely on the photometric properties of hot stars in
globular clusters. Much more physical information regarding their evolutionary
status can be gained from spectroscopic analyses: From spectra of various
resolutions in combination with photometric data we can determine their
atmospheric parameters (effective temperature, surface gravity, and helium
abundance) and compare those to the predictions of the stellar evolutionary
theory. The disadvantage of spectroscopic observations (compared to photometric 
ones) is the fact that they require larger telescopes and/or more observing
time: For the observations of the blue HB stars in M~15 we used the 3.5m
telescope of the German-Spanish Calar Alto observatory in Spain and the
targets in NGC~6752 were mostly observed with the NTT at the ESO La Silla
observatory in Chile.

To determine effective temperatures and surface gravities we compare various
spectroscopic and photometric observations to their theoretical counterparts.
Depending on the available observational and theoretical data and the amount
of software sophistication a wide variety of analysis methods is currently
available. The following paragraphs attempt to give an overview that allows to
judge our results -- for detailed information we refer the
reader to the cited papers.

\subsection{Effective Temperature}

\begin{figure}[t]
\vspace*{9cm}
\includegraphics{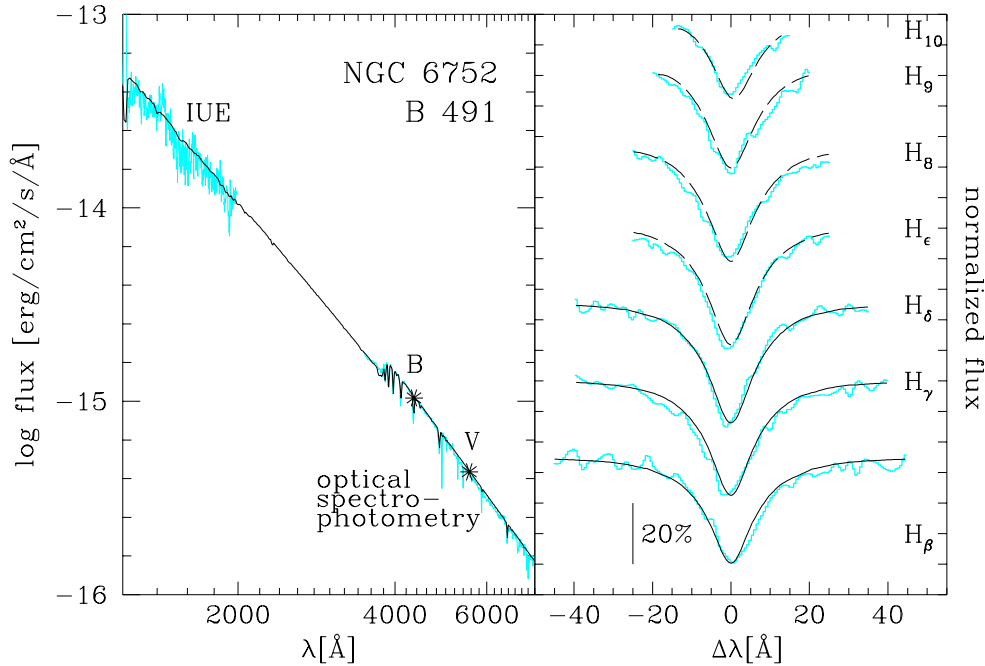}
\caption{The fit of the flux distribution ({\bf left}) and line profiles ({\bf
right}) of B~491 in NGC~6752. Plotted are the observed data (grey lines)
and a model atmosphere for [M/H] = $-$1.5, \teff\ = 28,500~K and \logg\ = 5.3. 
These parameters were determined from the flux distribution and the line 
profiles of H$_\beta$ to H$_\delta$ only. It can be seen that the 
corresponding model also fits the higher Balmer lines quite well.}
\label{plotfit}
\end{figure}

The ideal temperature indicator is insensitive to variations of surface
gravity because it then allows to derive the effective temperature 
independently from the surface gravity. The UV flux distribution meets this 
requirement for all blue HB stars and the Balmer jump fulfills it for stars 
with effective temperatures between about 11,000~K and 
30,000~K\footnote{Johnson UBV photometry becomes rather gravity independent
for temperatures above about 20,000~K - at the same time, however, it also
loses temperature sensitivity. Str\"omgren uvby photometry stays temperature
sensitive up to higher effective temperatures but is not available for most
globular clusters.}. As interstellar extinction changes (reddens) the flux
distribution of a star the observational data must be corrected for this
effect. We dereddened the observed spectra using the extinction law of Savage
\& Mathis (1979) and the appropriate reddening values for the respective
globular cluster. For the analysis published ATLAS9 model spectra (Kurucz
1992) for the metallicity closest to the globular cluster metallicity were
used. If not mentioned otherwise the quality of the fit was judged by eye. An
example is shown in Fig.~\ref{plotfit}. Temperature determinations that
include UV data (e.g. IUE spectrophotometry) are in general more reliable than
those relying solely on optical observations, as the UV flux distribution is
more sensitive to temperature variations than the optical continuum. However,
as only a very limited amount of UV spectrophotometry is available such data
were used only by Heber et al. (1986), de Boer et al. (1995), Cacciari et al.
(1995), and Moehler et al. (1997b).

If only {\bf optical spectrophotometry} is available (most stars in NGC~6752
and M~15) the overall continuum slope and the Balmer jump should be fitted
simultaneously, including - if possible - optical photometric data as well. It
turns out that straylight from red (i.e. cool) neighbours can cause problems
for optical spectrophotometry\footnote{This problem does not affect the IUE
data as the flux of cool stars decreases rapidly towards shorter
wavelengths.}: A model that fits the Balmer jump (and the BV photometry)
cannot fit the spectrophotometric continuum longward of 4000~\AA\ but instead
predicts too little flux there - the straylight from the cool star causes a
red excess. An attempt to quantify these effects is described in Moehler et
al. (1997b). 

Crocker et al. (1988) fit the continuum ($\lambda\lambda$ 3450--3700~\AA\ and
4000--5100~\AA ) in their spectrophotometric data and employ a $\chi^2$ test
to find the best fit. In addition they use the star's position along the
observed HB to obtain another estimate of its temperature and finally average
both values for \teff .

\subsection{Surface Gravity}

Provided the effective temperature has been determined as described above the
surface gravity can be derived by fitting the shape of the Balmer line 
profiles at a fixed temperature. For this purpose the spectra are normalized
and corrected for Doppler shifts introduced by the radial velocities of the
stars. The model spectra have to be convolved with a profile representing the
instrumental resolution, which was generally determined from the FWHM of the 
calibration lines (for more details see Moehler et al. 1995). We computed (at
fixed temperature) for H$_\beta$ to H$_\delta$ the squared difference between
the observed spectrum and the theoretical line profile and used the sum of
these differences as estimator for the quality of the fit (Moehler et al.
1995, 1997b, see also Fig.~\ref{plotfit}). Crocker et al. (1988) used the same
lines and employed a $\chi^2$ test to determine \logg . In addition they
corrected their results for the subsolar helium abundance normally present in
blue HB stars.

\subsection{Simultaneous Determination of \teff\ and \logg}

 For the cooler stars (below about 20,000~K) one can use a combination of {\bf
optical photometry} and {\bf Balmer line profile fits} to determine effective 
temperature and surface gravity simultaneously: Reddening free indices (Q for
Johnson UBV photometry, Moehler et al. 1995; [c1], [u-b] for Str\"omgren uvby
photometry, de Boer et al. 1995) in comparison with theoretical values allow
to determine a relation between effective temperature and surface gravity.
 Fits to the lower Balmer lines (H$_\beta$ to H$_\delta$) yield another relation
between \teff\ and \logg\ and from its intersection with the photometric
relation effective temperature and surface gravity can be derived (for
examples see de Boer et al. 1995 and Moehler \& Heber 1998).
 
 For stars below about 8,500~K (Moehler et al. 1995, M~15) the Balmer lines 
depend more on \teff\ than on \logg. In these cases the Balmer lines are used 
to estimate the temperature and \logg\ is derived from the Q value. 

Including also the higher Balmer lines (H$_\epsilon$ to H$_{10}$) allows to
derive effective temperature {\em and} surface gravity by fitting {\bf all
Balmer lines (H$_\beta$ to H$_{10}$)} simultaneously (Bergeron et al. 1992;
Saffer et al. 1994). This method has been used for the UV bright stars 
(Moehler et al. 1998a) and the blue HB stars in metal-rich globular clusters. 
We used the procedures developed by Bergeron et al. (1992) and Saffer et al. 
(1994), which employ a $\chi^2$ test to establish the best fit. Using only the
lower Balmer lines (H$_\beta$ to H$_\delta$) yields rather shallow minima of
$\chi^2$, which allow a large range of values for \teff\ and \logg .

\subsection{Helium Abundances}

Helium abundances were either derived from the simultaneous fitting of the 
Balmer and \ion{He}{i}/\ion{He}{ii} lines (Moehler et al. 1998a) or from
measured equivalent widths that are compared to theoretical curves-of-growth
for the appropriate values of effective temperature and surface gravity
(Moehler et al. 1997b).

\subsection{Model atmospheres}

Most of the stars discussed here are in a temperature--gravity range where LTE
(local thermal equilibrium) is a valid approximation for the calculation of
model atmospheres (Napiwotzki 1997). For the older data published ATLAS model
spectra were used: ATLAS6 (Kurucz 1979) by Crocker et al. (1988) resp. ATLAS9
(Kurucz 1992) by de Boer et al. (1995) and Moehler et al. (1995, 1997b). The
stars in NGC~6752 (Moehler et al. 1997b) required an extension of the model
atmosphere grid to higher surface gravities, for which we used an updated
version of the code of Heber (1983). The new fit procedures (Bergeron et al.
1992; Saffer et al. 1994) which we employed for the recent data (Moehler et
al. 1998a) required line profiles for the higher Balmer lines (shortward of
H$_\delta$) that are not available from Kurucz. We therefore calculated model
atmospheres using ATLAS9 (Kurucz 1991, priv. comm.) and used the LINFOR
program (developed originally by Holweger, Steffen, and Steenbock at Kiel
university) to compute a grid of theoretical spectra that contain the Balmer
lines H$_\alpha$ to H$_{22}$ and \ion{He}{i} lines. For those stars which show
\ion{He}{ii} lines in their spectra (and are thus considerably hotter than the
bulk of our programme stars) it is necessary to take non-LTE effects into
account (Napiwotzki 1997; Moehler et al. 1998a). 

\section{Gaps and Blue Tails}

\begin{figure}[t]
\vspace*{10.5cm}
\includegraphics{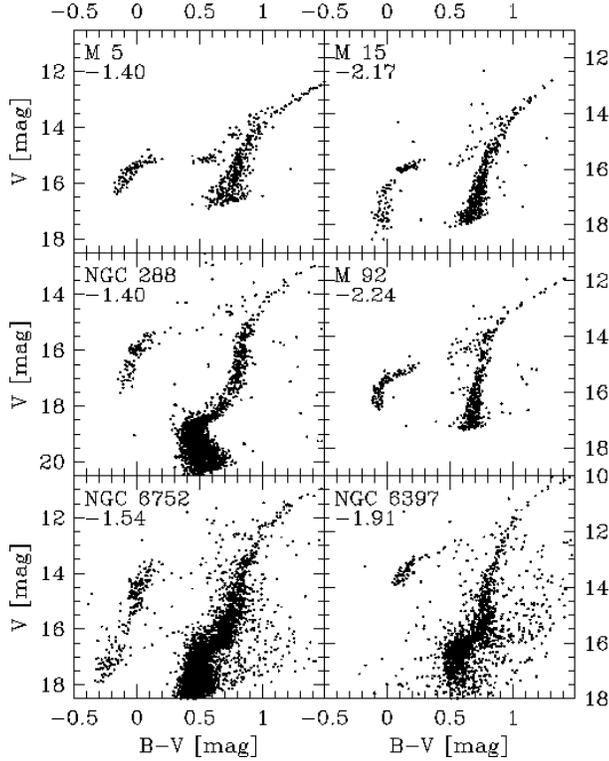}
\caption{Colour-magnitude diagrams of NGC~6752 (Buonanno et al. 1986), NGC~288
(Buonanno et al. 1984), M~5 (Buonanno et al. 1981), NGC~6397 (Alcaino et al.
1987), M~92 (Buonanno et al. 1983b), M~15 (Buonanno et al. 1983a). The wide
gaps in the brighter and more horizontal part of some HB's are in reality
populated by (variable) RR~Lyrae stars, which are omitted from the plots. The
gaps discussed in this paper are located at fainter magnitudes in the (mostly
vertical) blue tails (most pronounced in NGC~6752, NGC~288, and M~15). }
\label{cmds_obs}
\end{figure}

As mentioned above the blue tails seen in many CMD's of globular clusters are
often separated from the more horizontal part of the HB by gaps at varying
brightness (examples are shown in Fig.~\ref{cmds_obs}; for a list of globular
clusters with blue tails see Fusi Pecci et al. [1993]; Catelan et al. [1998]
and Ferraro et al. [1998] give comprehensive lists of clusters that show gaps
and/or bimodal horizontal branches). Such gaps can be found already in Arp's
(1955) CMD's and have caused a lot of puzzlement, since they are not
predicted by any canonical HB evolution. One of the first ideas was that the
gaps are created by diverging evolutionary paths that turn a unimodal
distribution on the ZAHB into a bimodal one as the stars evolve away from the
ZAHB (Newell 1973; Lee et al. 1994). Evolutionary calculations, however, do not
show any such behaviour as horizontal branch stars spend most of their
lifetime close to the ZAHB (Dorman et al. 1991; Catelan et al. 1998). Rood \& 
Crocker (1985) suggested that the gaps separate two groups of HB stars that 
differ in, e.g., CNO abundance or core rotation. A more extreme version of
this idea was suggested by Iben (1990): blue tail stars are produced
differently from the blue HB stars, e.g. by merging of two helium white
dwarfs. So far, no precursor systems of such stars have been observed. Quite
recently, Caloi (1999) proposed a change in the stellar atmospheres from
convection to diffusion as an explanation for the gaps around (B$-$V)$_0$ = 0,
whereas Catelan et al. (1998) suggested that at least some of the gaps may be
due to statistical fluctuations. More detailed descriptions of possible 
explanations for the gaps can be found in Crocker et al. (1988), Catelan et 
al. (1998), and Ferraro et al. (1998).

The need for more information on the stars along the blue tails led to our
project to obtain atmospheric parameters for blue HB and blue tail stars in
several globular clusters: NGC~6397 (de Boer et al. 1995), NGC~6752 (Heber et
al. 1986; Moehler et al. 1997b), and M~15 (Moehler et al. 1995, 1997a). To
enlarge our sample we also used the data of NGC~288, M~5, and M~92 published
by Crocker et al. (1988). The CMD's of these clusters can be found in 
 Fig.~\ref{cmds_obs}.

\subsection*{Evolutionary status}

\begin{figure}
\vspace*{8.6cm}
\includegraphics{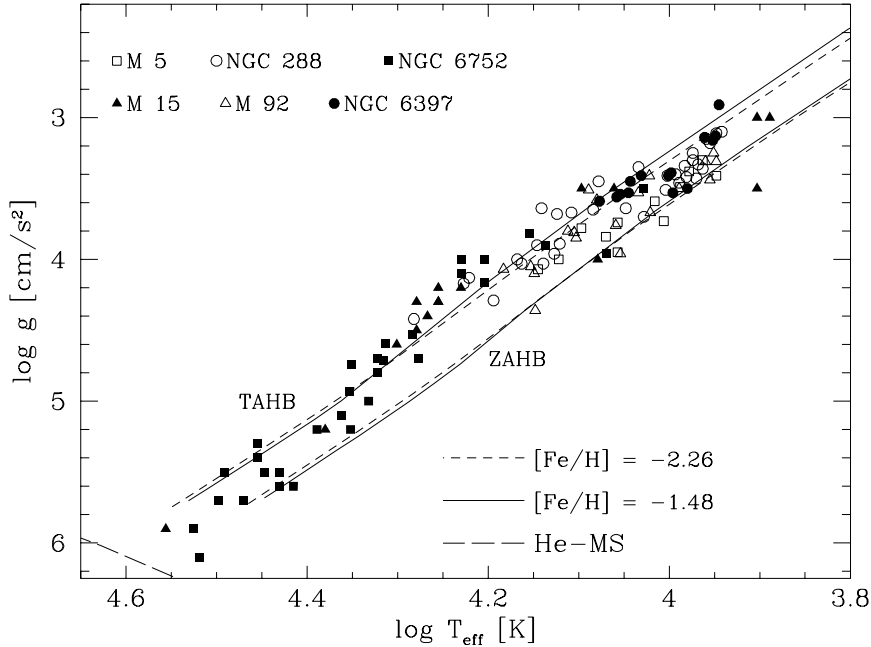}
\caption{The results of Crocker et al. (1988; M~5, M~92, NGC~288), de Boer 
et al. (1995; NGC~6397) and Moehler et al. (1995, 1997a, M~15; 1997b, NGC6752)
compared to evolutionary tracks from Dorman et al. (1993). ZAHB and TAHB 
stand for zero-age and terminal-age HB (see text for details). Also given is 
the position of the helium main-sequence (Paczynski 1971).}
\label{plottga1}
\end{figure}

In Fig.~\ref{plottga1} the physical parameters of the HB stars analysed by
Crocker et al. (1988; M~5, M~92, NGC~288), de Boer et al. (1995; NGC~6397), and
Moehler et al. (1995, 1997a, M~15; 1997b, NGC6752) are compared to 
evolutionary tracks. The zero-age HB (ZAHB) marks the position where the HB 
stars have settled down and started to quietly burn helium in their cores. The
terminal-age HB (TAHB) is defined by helium exhaustion in the core of the HB
star. The distribution of stars belonging to an individual cluster is hard to
judge in this plot but it is obvious that the observed positions in the (\logg
, \teff)-diagram fall mostly above the ZAHB and in some cases even above the
TAHB\footnote{Preliminary results of Bragaglia et al. (1999) indicate
deviations from this trend}. An indication of a low-temperature gap can be seen
at \logt\ $\approx$ 4.1. The gaps seen in the CMD of NGC~6752 and in the M~15
data of Durrell \& Harris 1993 (from which the two hottest stars in M~15 were
selected) are visible in the (\logg , \teff) plane at about 24,000~K, where
they separate BHB from EHB stars. In all other clusters the stars above and
below the gaps are blue horizontal branch stars cooler than 20,000~K.

Independent of the occurrence of any gaps stars with temperatures between 
11,000 (\logt\ = 4.04) and 20,000~K (\logt\ = 4.30) show lower gravities than 
expected from canonical scenarios, whereas stars outside this temperature 
range are well described by canonical HB and EHB evolutionary tracks. The UIT
observations of M~13 (Parise et al. 1998) and the HUT spectra of M~79 (Dixon
et al. 1996) also suggest lower than expected gravities for blue HB stars.
Whitney et al. (1998) use UIT observations of the hot stars in $\omega$ Cen to
claim that the extreme HB stars -- which agree with theoretical expectations
in our results -- have lower than expected luminosities, which would mean
higher than expected gravities. These deviations in \logg\ could indicate that
some assumptions used for the calculations of model atmospheres may not be 
appropriate for the analysis of the BHB stars (see also de Boer et al. 1995, 
Moehler et al. 1995):

Diffusion might lead to peculiar abundance patterns, because radiative 
levitation might push up some metals into the atmospheres whereas other
elements might be depleted due to gravitational settling. Line blanketing
effects of enhanced metals may change the atmospheric structure. We found,
however, that even an increase of 2 dex in [M/H] results in an increase of
only 0.1~dex in \logg\footnote{This is consistent with the findings of Leone \&
Manfr\`e (1997) that Balmer-line gravities can be underestimated by 0.25 dex
if a solar metal abundance is assumed for metal-rich helium weak stars.}. 
Another effect of diffusion might be a stratification of the atmosphere, i.e.
an increase of helium abundance with depth, which has been predicted for white
dwarf atmospheres (Jordan \& Koester, 1986). In order to affect the Balmer
jump significantly the transition from low to high He abundance must take place
at an optical depth intermediate between the formation depths of the Paschen
and the Balmer continua. Such a fine tuning is unlikely to occur.

Rapid rotation rotation of the stars -- if neglected in the model atmospheres 
-- would lower the determined gravities. This effect, however, becomes 
significant only if the rotation velocity exceeds about half of the break-up
velocity. As measured rotation velocities for HB stars are small (Peterson et
al. 1995) this possibility can be ruled out as well.

As we did not find any systematic effects in our analysis that are large
enough to explain the observed offsets in surface gravity we assume for now
that the physical parameters are correct and look for a scenario that can
explain them\footnote{Scenarios like the merging of two helium white dwarfs
(Iben \& Tutukov 1984) or the stripping of red giant cores (Iben \& Tutukov
1993, Tuchman 1985) may produce stars that deviate from the ZAHB. Such stars,
however, are either too hot (merger) or too short-lived (stripped core) to
reproduce our results.}:

\subsubsection*{Deep mixing}

Abundance variations (C, N, O, Na, Al) in globular cluster red giant stars
(Kraft 1994, Kraft et al. 1995, Pilachowski et al. 1996) suggest that
nuclearly processed material from deeper regions is mixed to the surface of
the stars. Depending on the element considered this mixing has to reach down
into varying depths. The enhancement of aluminium, for instance, requires the
mixing to extend down into the hydrogen burning (= helium producing) shell
(e.g. Cavallo et al. 1998). This means that any mixing that dredges up
aluminium will also dredge up helium ({\it helium mixing} or {\it deep
mixing}). Table~\ref{tab-mix} lists the evidence for deep mixing for the
clusters shown in Fig.~\ref{cmds_obs}.

\begin{table}
\begin{tabular}{lll}
Cluster & Mixing & Reference\\
\hline
NGC 6752 & probable & Shetrone 1997, IAU Symp. 189(P), 158\\
NGC 6397 & probable & Bell et al. 1992, AJ 104, 1127\\
M 92     & probable & Shetrone 1996, AJ 112, 1517\\
M 15     & probable & Sneden et al. 1997, AJ 114 1964\\
M 5      & unlikely & Sneden et al. 1992, AJ 104, 2121\\
NGC 288  & unlikely & Dickens et al. 1991, Nature 351, 212\\
\hline
\end{tabular}
\caption[]{References concerning deep mixing in the clusters shown in 
 Fig.~\ref{cmds_obs}}
\label{tab-mix}
\end{table}

\suppressfloats

Such ``helium mixed'' red giant stars evolve to higher luminosities and 
therefore lose more mass than their canonical counterparts. The resulting HB
stars then have less massive hydrogen envelopes and are thus hotter than in the
canonical case. In addition the higher helium abundance in the hydrogen
envelopes of the HB stars increases the efficiency of the hydrogen shell
burning and thereby leads to higher luminosities at a given effective
temperature. This increase in luminosity leads to lower gravities for ``deep
mixed'' HB stars than predicted by canonical evolution. For a more detailed
discussion of the effects of deep mixing on post-RGB evolution see Sweigart
(1997, 1999).

\begin{figure}[!]
\vspace*{8.6cm}
\includegraphics{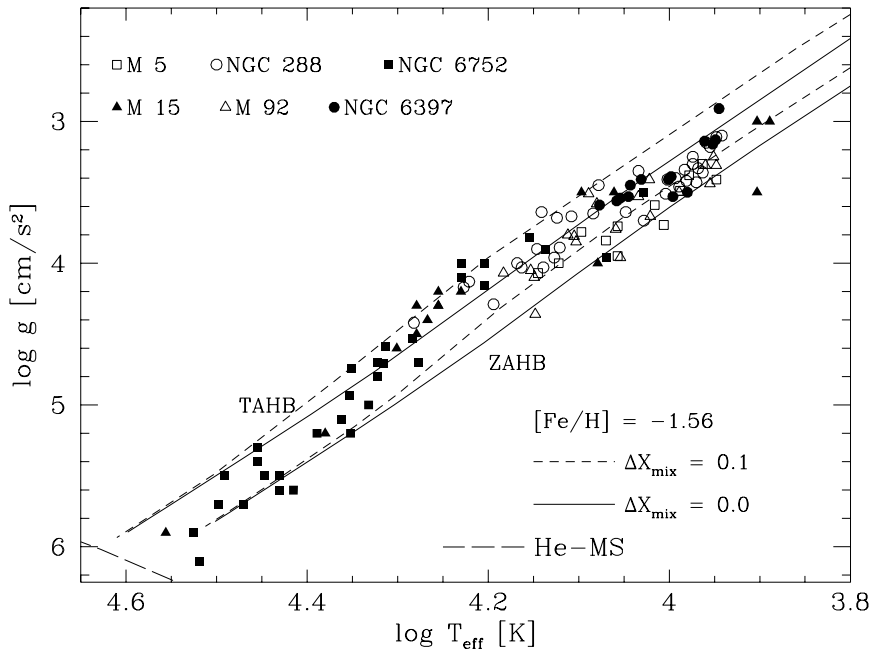}
\caption{The results from Fig.~\ref{plottga1} compared to evolutionary tracks
that take into account the effects of helium mixing (Sweigart 1999). $\Delta
X_{mix}$ gives the amount of He mixing with 0 indicating an unmixed track (for
details see Sweigart 1999). Also given is the position of the helium
main-sequence (Paczynski 1971).}
\label{plottga2}
\end{figure}

 From Fig.~\ref{plottga2} it can be seen that most stars hotter than 11,000~K 
are well fitted by tracks that assume deep mixing\footnote{The good fit of the
helium-mixed tracks to the stars in NGC~288 is problematic as there is no
evidence for deep mixing in this cluster (cf. Table~\ref{tab-mix}).}. The
cooler stars, however, are better explained by canonical evolution. As deep
mixing leads to hotter and brighter blue HB stars it is possible that cool
blue HB stars result from unmixed RGB stars. Unfortunately it is not possible
to determine the envelope helium abundance of a blue HB star because almost
all of these stars are helium-deficient due to diffusion (Heber 1987). As it
remains unclear what causes deep mixing (although rotation probably plays a
role, Sweigart \& Mengel 1979) we do also not know whether all RGB stars
within one cluster experience the same degree of mixing. 

\subsection*{Masses}

Knowing effective temperatures and surface gravities of the stars allows to
determine the theoretical brightness at the stellar surface, which together
with the absolute brightness of the star yields its radius and thus its mass 
(see de Boer et al. 1995, Moehler et al. 1995, 1997b). The distances to the
globular clusters (necessary to determine the absolute brightnesses of the
stars) were taken from the compilation of Djorgovski (1993). The results are
plotted in Fig.~\ref{plottlm1} and can be summarized as follows:

\begin{figure}[!]
\vspace*{8.6cm}
\includegraphics{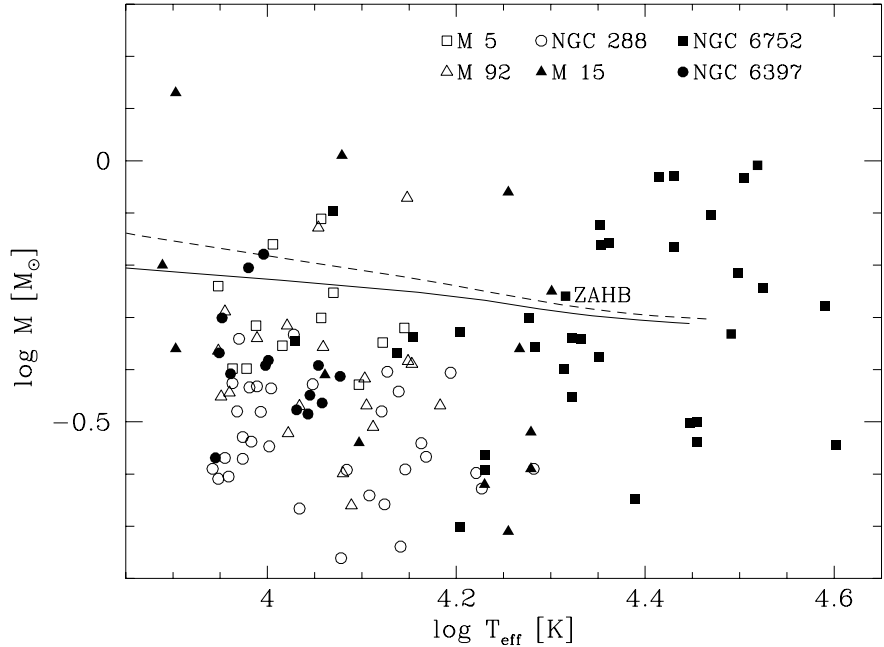}
\caption{The masses of the stars derived with the distances listed by
Djorgovski (1993) compared to evolutionary calculations from Dorman et al.
(1993). The solid line gives the ZAHB for [Fe/H] = $-$1.48, the dashed line
marks the ZAHB for [Fe/H] = $-$2.26.}
\label{plottlm1}
\end{figure}

While the masses of the stars in M~5 and NGC~6752 scatter around the canonical
values the blue HB stars in all other clusters show masses that are
significantly lower than predicted by canonical HB evolution - even for
temperatures cooler than 11,000~K where we saw no deviation in surface gravity
from the canonical tracks. The fact that the stars in two of the clusters show
``normal'' mass values makes errors in the analyses an unlikely cause for the
problem (for a more detailed discussion see Moehler et al. 1995, 1997b, and de
Boer et al. 1995). Also the merger models of Iben (1990) cannot explain these 
masses since the resulting stars are much hotter. However, if some of the {\bf
distance moduli} we used were too small this could cause such an effect --
larger distances would result in brighter absolute magnitudes, i.e. larger
radii and thus larger masses. 

\subsubsection*{Distances to Globular Clusters }

Using {\sc Hipparcos} data for local subdwarfs several authors (Reid 1997, 
1998; Gratton et al. 1997; Pont et al. 1998) determined distances to globular
clusters by main sequence fitting. The results are given in
Table~\ref{tab-hipp} and show that the new distance moduli are in general
larger than the old ones, in some cases by up to \magpt{0}{4} \ldots
\magpt{0}{6}\footnote{An increase of \magpt{0}{2} in (m-M)$_{\rm V}$
increases the mass of a cluster star by 20\%.}. 

\begin{table}
\begin{tabular}{llllllll}
\hline
Cluster & \multicolumn{2}{c}{[Fe/H]} & \multicolumn{5}{c}{(m-M)$_0$}\\
 & ZW84 & CG97 & D93 & R97 & R98 & G97 & P98 \\
\hline
47 Tuc         & $-$0.71 & $-$0.67 & 13.31 & & 13.56 & 13.64 & \\
M 71           & $-$0.58 & $-$0.70 & 12.96 & & 13.19 & & \\
{\bf NGC 288}  & $-$1.40 & $-$1.05 & 14.62 & & 15.00 & 14.96 & \\
{\bf M 5}      & $-$1.40 & $-$1.10 & 14.40 & 14.45 & & 14.62 & \\
NGC 362        & $-$1.28 & $-$1.12 & 14.67 & & & 15.06 & \\
M 13           & $-$1.65 & $-$1.41 & 14.29 & 14.48 & 14.45 & 14.47 & \\
{\bf NGC 6752} & $-$1.54 & $-$1.43 & 13.12 & 13.20 & 13.16 & 13.34 & \\
{\bf NGC 6397} & $-$1.91 & $-$1.82 & 11.71 & & 12.24 & & \\
M 30           & $-$2.13 & $-$1.88 & 14.35 & 14.95 & & 14.96 & \\
M 68           & $-$2.09 & $-$1.95 & 14.84 & 15.29 & & 15.33 & \\
{\bf M 15}     & $-$2.15 & $-$2.12 & 15.11 & 15.38 & & & \\
{\bf M 92}     & $-$2.24 & $-$2.15 & 14.38 & 14.93 & & 14.82 & 14.70 \\
\hline
\end{tabular}
\caption[]{{\sc Hipparcos} based distance moduli for globular clusters 
compared to older values. The data are taken from Zinn \& West (1984, ZW84),
Carretta \& Gratton (1997, CG97), Djorgovski (1993, D93), Reid (1997, R97; 
1998, R98), Gratton et al. (1997, G97) and Pont et al. (1998, P98). The 
clusters discussed here are marked in bold font.}
\label{tab-hipp}
\end{table}

\begin{table}[!]
\begin{tabular}{lllll}
\hline
Cluster & \multicolumn{2}{c}{[Fe/H]} & $<\frac{M}{M_{ZAHB}}>_{D93}$ 
 & $<\frac{M}{M_{ZAHB}}>_{R}$ \\
\hline
NGC 288  & $-$1.40 & $-$1.05 & 0.52 $\pm$ 0.12 & 0.69 $\pm$ 0.16 \\
M 5      & $-$1.40 & $-$1.10 & 0.87 $\pm$ 0.21 & 0.95 $\pm$ 0.22 \\
NGC 6752 & $-$1.54 & $-$1.43 & 1.04 $\pm$ 0.52 & 1.10 $\pm$ 0.55 \\
NGC 6397 & $-$1.91 & $-$1.82 & 0.62 $\pm$ 0.16 & 0.78 $\pm$ 0.20 \\
M 15     & $-$2.15 & $-$2.12 & 0.81 $\pm$ 0.49 & 1.10 $\pm$ 0.67 \\
M 92     & $-$2.24 & $-$2.15 & 0.66 $\pm$ 0.25 & 1.10 $\pm$ 0.42 \\
\hline
\end{tabular}
\caption[]{The average ratio of calculated mass (as described in the text) 
to the mass on the ZAHB (for the respective temperature) of HB stars in 
globular clusters plotted in Figs.~\ref{plottlm1} and \ref{plottlm2}. The 
distance moduli were taken from Djorgovski (1993, D93) and Reid 
(1997, 1998, R).}
\label{tab-mass}
\end{table}

\begin{figure}[!]
\vspace*{8.6cm}
\includegraphics{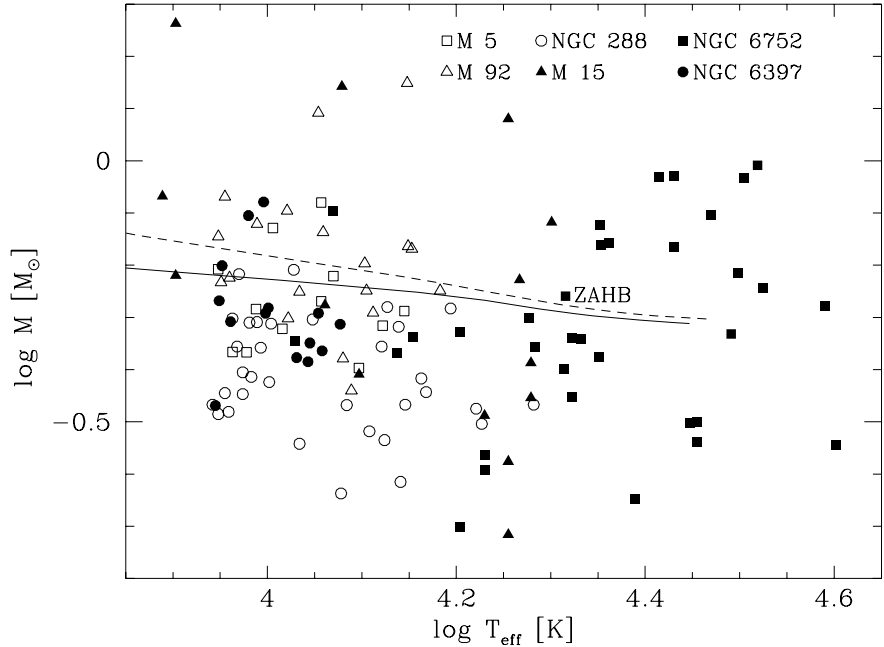}
\caption{The masses of the stars derived with the new distances listed by Reid
(1997, 1998) compared to evolutionary tracks from Dorman et al. (1993). The
solid line gives the ZAHB for [Fe/H] = $-$1.48, the dashed line marks the ZAHB
for [Fe/H] = $-$2.26.}
\label{plottlm2}
\end{figure}

It is interesting to note that for M~5 and NGC~6752 (where the masses almost
agree with the canonical expectations) the new distances are close to the old
ones, whereas for the metal poor clusters M~15, M~92, and NGC~6397 the new 
distance moduli are \magpt{0}{3} -- \magpt{0}{6} larger than the old ones, 
thereby greatly reducing the mass discrepancies (see also Heber et al., 1997). 
The resulting new masses are plotted in Fig.~\ref{plottlm2} and in
Table~\ref{tab-mass} we list the average ratio between the mass calculated for
an HB star (as described in the text) and the supposed ZAHB mass for its
temperature (from Dorman et al. 1993). It can be seen that in all cases 
(except NGC~6752) the agreement between expected and calculated mass improves
with the new distance moduli, although the masses in NGC~288 remain
significantly too low. From our observations we therefore favour the longer
distance scale for globular clusters as suggested by most analyses of the {\sc
Hipparcos} data.

\section{Blue HB Stars in Metal-Rich Globular Clusters }

As mentioned in Section~1 metal-rich globular clusters tend to have red
horizontal branches. This is plausible as according to canonical stellar 
evolutionary theory metal-rich HB stars have to have much smaller envelope 
masses than metal-poor HB stars to achieve the same temperature. Therefore a
fine tuning of mass loss is required to produce blue HB stars in metal-rich 
environments in the framework of classical stellar evolution. Deep mixing or
the merging of two helium white dwarfs offer other, more exotic, possibilities
to produce hot stars. Yi et al. (1997, 1998) discuss possible mechanisms to
produce blue HB stars in elliptical galaxies and d'Cruz et al. (1996) describe
mechanisms to create extreme HB stars in metal-rich open clusters like
NGC~6791, where Liebert et al. (1994) found subdwarf B stars. Despite this
recent theoretical work it came as a surprise when blue HB stars really showed
up in metal-rich globular clusters:

UIT images of 47~Tuc ([Fe/H] = $-$0.71; O'Connell et al. 1997) and of
NGC~362\footnote{\rm While not exactly metal-rich NGC~362 has been famous as
part of the second-parameter pair NGC~288/NGC~362: Both clusters have similar
metallicities, but NGC~288 shows a well populated blue HB, whereas NGC~362
shows almost only red HB stars.} ([Fe/H] = $-$1.28; Dorman et al. 1997) show
the presence of blue stars. Colour-magnitude diagrams of the central regions of
NGC~6388 ([Fe/H] = $-$0.60) and NGC~6441 ([Fe/H] = $-$0.53) obtained with the
Hubble Space Telescope (HST) show sloped blue HB's and long blue tails in
both clusters (Rich et al. 1997). The slope means that in these clusters bluer
HB stars are visually brighter than redder ones, in contrast to canonical 
expectations. Their brighter luminosities require the blue HB stars to have
lower gravities, which can be caused by rotation, deep mixing and/or higher
primordial helium abundance (Sweigart \& Catelan 1998).

\begin{figure}[!]
\vspace*{8.6cm}
\includegraphics{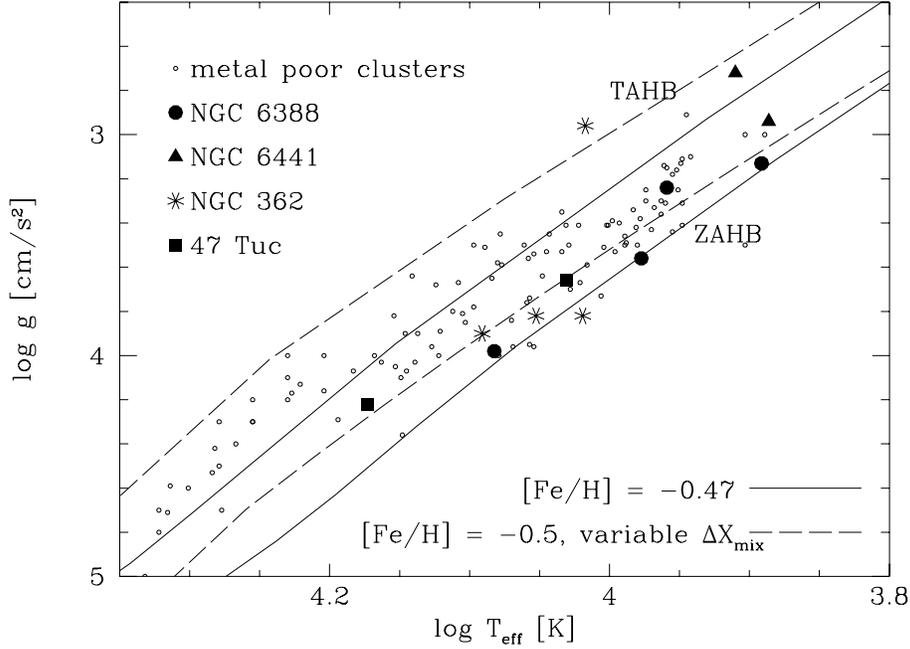}
\caption{The effective temperatures and surface gravities of the blue stars in
47~Tuc, NGC~362, NGC~6388, and NGC~6441 compared to an evolutionary track for
[Fe/H] = $-$0.47 from Dorman et al. (1993, solid line). Only probable cluster
members are plotted. The dashed lines mark the region of deep mixed tracks for
[Fe/H] = $-$0.5 of Sweigart (priv. comm.).}
\label{plottg3}
\end{figure}

To find out what really causes the unexpected presence of blue HB stars in 
these metal-rich clusters we decided to perform a spectroscopic investigation. 
Unfortunately we were not too lucky with weather and technical conditions and
the number of observed stars is small. In addition, some of the blue stars in
47~Tuc and NGC~362 turned out to be field HB stars or SMC main sequence stars.
Those stars, that are members of the clusters, are confirmed to be blue HB
stars with effective temperatures between 7,500~K and 15,000~K. Due to the
weather conditions we could not observe the fainter and therefore hotter stars
in these clusters, which would be especially interesting for the question of
deep mixing. More details can be found in the forthcoming papers Moehler,
Landsman, Dorman (47~Tuc, NGC~362) and Moehler, Catelan, Sweigart, Ortolani
(NGC~6388, NGC~6441). The results of our spectroscopic analyses are plotted in
 Fig.~\ref{plottg3} and show that so far there is no evidence for deep mixing or
primordial helium enrichment in these stars\footnote{The helium white dwarf
merging model of Iben (1990) is unable to produce stars with so low
temperatures, because available hydrogen envelope masses are small ($<
10^{-4}$~\Msolar).}. More spectra, especially of fainter stars, are necessary
to verify this statement.

\section{Hot UV Bright Stars in Globular Clusters}

\begin{figure}[ht]
\vspace*{8.6cm}
\includegraphics{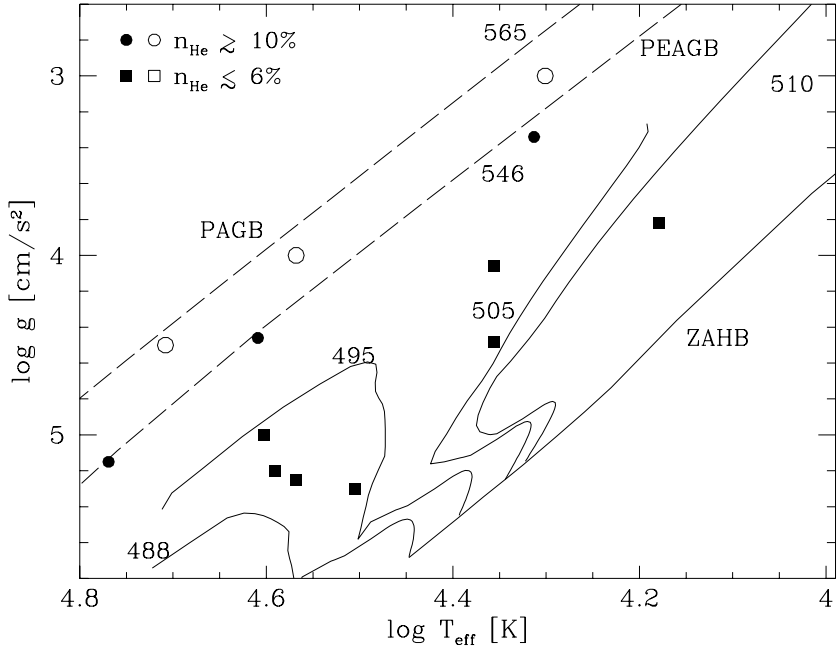}
\caption{The atmospheric parameters of the UV bright stars compared to 
evolutionary tracks. The solid lines mark the ZAHB and post-EHB evolutionary 
tracks for [Fe/H] = $-$1.48 (Dorman et al. 1993). The dashed lines give 
post-AGB (0.565~\Msolar) and post-early AGB (0.546~\Msolar) tracks from 
Sch\"onberner (1983). All evolutionary tracks are labeled with the mass of the 
respective model in units of 10$^{-3}$\Msolar. The filled symbols are from
Moehler et al. (1997b, 1998a), the open symbols are taken from Conlon et al.
(1994), Heber \& Kudritzki (1986) and Heber et al. (1993).}
\label{uvbs_tg}
\end{figure}

As mentioned in Section~1 optical searches for UV bright stars in globular 
clusters yielded mainly stars cooler than 30,000~K (the majority of which was
even cooler than 15,000~K) due to the increasing bolometric corrections for
hotter stars. The vast majority of stars selected this way will evolve either
from the blue HB to the asymptotic giant branch or from there to the white
dwarf domain. It is rather unlikely to find post-EHB stars this way as they
spend only short time in such cool regions (if they reach them at all, see
 Fig.~\ref{uvbs_tg}). In addition their overall fainter magnitudes work against
their detection. Searches in the ultraviolet regime, on the other hand, will
favour hotter stars and thereby increase the chance to detect post-EHB stars.
We therefore decided to spectroscopically analyse the many hot UV bright stars
that were found in globular clusters by the Ultraviolet Imaging Telescope.
Details of the observations, reduction, and analyses can be found in Moehler 
et al. (1998a). The main goal was to find out how the physical parameters of
these stars compare to evolutionary tracks.

The derived effective temperatures and gravities of the target stars are
plotted in Fig.~\ref{uvbs_tg} and compared to various evolutionary tracks.
Three of the stars (in NGC~6121 and NGC~6723) appear to fit the post-early 
AGB\footnote{Post-early AGB stars left the asymptotic giant branch before the 
thermally pulsing stage} track, while the remaining targets (in NGC~2808 and
NGC~6752\footnote{including three stars analysed by Moehler et al. (1997b)})
are consistent with post-EHB evolutionary tracks. Like the extreme HB stars
themselves (Moehler et al. 1997b) the post-EHB stars show subsolar helium
abundances probably caused by diffusion. In contrast the three post-early AGB
stars - which are supposed to be successors to helium-deficient blue HB stars
- have approximately solar helium abundances. This agrees with the expectation
that already during the early AGB stages convection is strong enough to
eliminate any previous abundance patterns caused by diffusion.

As expected, the two clusters with populous EHB's (NGC~2808 and NGC~6752) have
post-EHB stars but no post-AGB stars. The number ratio of post-EHB to EHB 
stars in NGC~6752, however, is much lower than expected from stellar 
evolutionary theory: 6\% instead of 15 -- 20\%. This discrepancy has first
been noted by Landsman et al. (1996) and has been confirmed by our studies,
which verified all four post-EHB candidates, but found no additional ones 
(Moehler et al. 1997b, 1998a). The clusters NGC~6723 and M~4, on the other
hand, do not have an EHB population, although they do have stars blueward of
the RR Lyrae gap (which are potential progenitors of post-early AGB stars). 
The lack of genuine post-AGB stars may be understood from the different
lifetimes: The lifetime of Sch\"onberner's (1983) post-early AGB track is
about 10 times longer than that of his lowest mass post-AGB track. Thus, even
if only a small fraction of stars follow post-early AGB tracks, those stars
may be more numerous than genuine post-AGB stars. Due to their relatively long
lifetime, post-early AGB stars are unlikely to be observed as central stars of
planetary nebulae since any nebulosity is probably dispersed before the
central star is hot enough to ionize it. These different life times in 
combination with the fact that a considerable number of globular clusters 
stars (all post-EHB stars and some post-BHB stars) do not reach the thermally
pulsing AGB stage could be an explanation for the lack of planetary nebulae in
globular clusters reported by Jacoby et al. (1997). 

\section{Abundance Patterns of UV Bright Stars in Globular Clusters}

Up to now detailed abundance analyses have been performed mainly for post-AGB
stars in the field of the Milky Way (McCausland et al. 1992 and references
therein; Conlon 1994; Napiwotzki et al. 1994; Moehler \& Heber 1998), for
which the population membership is difficult to establish. The summarized
result of these analyses is that the abundances of N, O, and Si are roughly
1/10 of the solar values, while Fe and C are closer to 1/100 solar.
McCausland et al. (1992) and Conlon (1994) interpret the observed abundances
as the results of dredge-up processes on the AGB, i.e. the mixing of
nuclearly processed material from the stellar interior to the surface.
Standard stellar evolutionary theories (Renzini \& Voli 1981; Vassiliadis \&
Wood 1993) do not predict any dredge-up processes for the low-mass precursors
of these objects. Nevertheless the planetary nebula Ps~1 in M~15 as well as the 
atmosphere of its central star K~648 are both strongly enriched in carbon when
compared to the cluster carbon abundance\footnote{A preliminary analysis of
ZNG1 in M~5 also shows evidence for a third dredge-up, but no trace of a nebula 
(Heber \& Napiwotzki 1999).} (Adams et al. 1984; Heber et al. 1993),
pinpointing the dredge-up of triple $\alpha$ processed material to the stellar
surface and suggesting a possible connection between dredge-up and planetary
nebula ejection (Sweigart 1998). This discrepancy may be solved by newer
evolutionary calculations which are able to produce a third dredge-up also in
low-mass AGB stars (Herwig et al. 1997).

Napiwotzki et al. (1994) on the other hand suggest that the photospheric
abundances are caused by gas-dust separation towards the end of the AGB phase:
If the mass loss at the end of AGB ceases rapidly gas can fall back onto the 
stellar surface while the dust particles are blown away by radiative pressure. 
This process has been proposed by Bond (1991) to explain the extreme iron 
deficiencies seen in some cooler post-AGB stars and is described in more 
detail by Mathis \& Lamers (1992).

As iron is very sensitive to depletion by gas-dust separation the iron
abundance is the crucial key to the distinction between dredge-up and gas-dust
separation. To verify any elemental depletion, however, one has to know the
original abundance of the star, which is generally not the case for field
stars. Therefore UV bright stars in globular clusters with known metallicities
provide ideal test cases for this problem and we started a project to derive
iron abundances from high-resolution UV spectra obtained with HST.

\subsection*{Abundance analysis of Barnard~29 in M~13 and ROA~5701 in 
$\omega$ Cen}

\begin{figure}[!ht]
\vspace{10cm}
\includegraphics{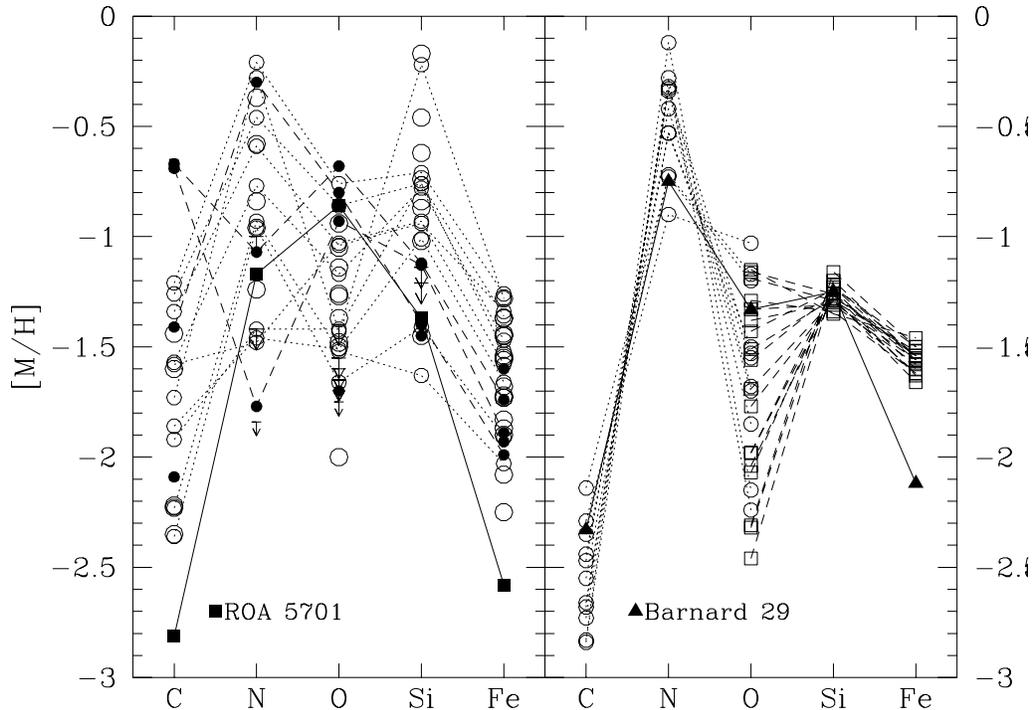}
\caption{The abundances derived for ROA~5701 ({\bf left}) and Barnard~29 ({\bf
right}) compared to those of other stars in $\omega$ Cen resp. M~13. The solid
line connects the abundances derived for ROA~5701 resp. Barnard~29. In the
{\bf left} panel the short dashed lines mark the cool UV bright stars in
$\omega$ Cen (filled circles, Gonzalez \& Wallerstein 1994), the dotted lines
connect the abundances of the red giants (open circles, Paltoglou \& Norris
1989; Brown et al. 1991; Brown \& Wallerstein 1993; Smith et al. 1995). In
the {\bf right} panel the dotted and short-dashed lines mark the abundances of
the red giants in M~13 taken from Smith et al. (1996, circles) resp. Kraft et
al. (1997, squares).}
\label{ghrs_uvbs}
\end{figure}

 For Barnard~29 a detailed abundance analysis from optical spectra has been 
done by Conlon et al. (1994) and we use their results for \teff\ and \logg\ 
for our analysis. For ROA~5701 we determined these parameters from IUE
low-resolution spectra, optical photometry, and optical spectroscopy. For the
iron abundances of both stars we used GHRS spectra of 0.07~\AA\ resolution
that cover the range 1860 $-$ 1906 \AA. The abundances have been derived using
the classical curve-of-growth technique. We computed model atmospheres for the
appropriate values of effective temperature, surface gravity, and cluster
metallicity and used the LINFOR spectrum synthesis package (developed
originally by Holweger, Steffen, and Steenbock at Kiel university) for the
further analysis. A more detailed description of our analysis can be found in
Moehler et al. (1998b).

 For ROA~5701 we find an iron abundance of $\log{\epsilon_{Fe}} = 4.89 \pm 
0.12$ ([Fe/H] = $-$2.61) and for Barnard~29 we get $\log{\epsilon_{Fe}} = 5.38
\pm 0.14$ ([Fe/H] = $-$2.12). Both stars thus show iron abundances 
significantly below the mean cluster abundances of [Fe/H] $\approx$ $-$1.5
\ldots $-$1.7. To look for any abundance trends in Barnard~29 and ROA~5701 in
comparison with other stars in these clusters we used the abundances of C, N,
O, Si in addition to iron. For ROA~5701 we determined these abundances from
optical high-resolution spectra. Abundance analyses of Barnard~29 have been
performed by Conlon et al. (1994, N, O, Si) and Dixon \& Hurwitz\footnote{They
also give an iron abundance of $\log{\epsilon_{Fe}} = 5.30^{+0.22}_{-0.26}$,
somewhat higher than ours.} (1998, C). 

\begin{figure}[ht]
\vspace{10cm}
\includegraphics{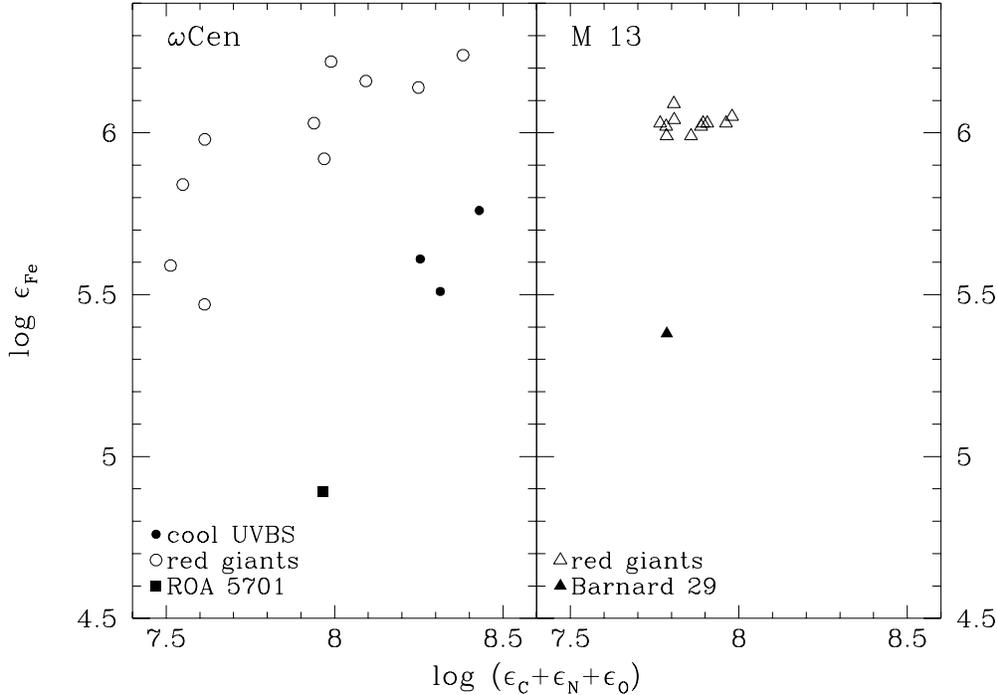}
\caption[]{The Fe abundances vs. the sum of CNO abundances for ROA~5701 ({\bf
left panel}) and Barnard~29 ({\bf right panel}) compared to those of red
giants and cool UVBS in $\omega$ Cen and M~13, respectively. For references
see Fig.~\ref{ghrs_uvbs}.}
\label{ghrs_cno}
\end{figure}

It can be seen in Fig.~\ref{ghrs_uvbs} that N, O, and Si in our two objects
show a behaviour similar to that in red giant stars. C seems to be depleted in 
ROA~5701, but this abundance is based on an upper limit for one line only
(4267~\AA) which may be affected by non-LTE effects. One should note here,
however, that Gonzalez \& Wallerstein (1994) find strong enhancements of CNO
and s-process elements for the brightest of the cool UV bright stars in
$\omega$~Cen which they interpret as evidence for a third dredge-up. Taking
the sum of C+N+O as indicator for the original iron abundance (cf.
 Fig.~\ref{ghrs_cno}) shows that ROA~5701 and Barnard~29 were not born iron
depleted and -- contrary to the brightest cool UV bright stars in $\omega$ Cen
-- also do not show any evidence for a third dredge up. The results of our
analysis thus favour the gas-dust separation scenario as explanation for the
abundance patterns of low-mass post-AGB and post-early AGB stars.

\section{Summary}

About nine years ago we began a spectroscopic study of blue horizontal branch
stars to find the reason for the gaps seen in the CMD's of many globular
clusters. While we haven't yet achieved this goal the study led to others
which altogether resulted in some interesting findings about the evolutionary
status of hot stars in globular clusters:

\subsection*{Blue Horizontal Branch Stars}

We studied stars above and below the gaps seen along the blue horizontal branch 
in the CMD's of many globular clusters (cf. Fig.~\ref{cmds_obs}) and found that
most of the stars below the gaps are physically the same as the stars above
the gaps, i.e. blue horizontal branch stars with a helium burning core and a
hydrogen burning shell. So far extreme horizontal branch stars have been
verified spectroscopically only in two clusters (NGC~6752 and M~15). 

The blue HB stars with temperatures between 11,000~K and 20,000~K show lower
gravities than expected from canonical stellar evolution, which can be
explained by deep mixing. The lower than expected masses that are found for
most stars cooler than 20,000~K can be understood if we assume that the
distance moduli to the globular clusters are larger than previously thought. 
Analyses of BHB and EHB stars within the same cluster will provide a crucial
test of these two hypotheses.

We verified that most of the blue stars seen in the colour-magnitude diagrams 
of several metal-rich globular clusters are indeed blue horizontal branch 
stars in these clusters. We did so far not find any significant evidence for
deep mixing or a higher primordial helium abundance in these metal-rich
globular clusters but have currently too few data to draw any firm conclusions. 

\subsection*{UV Bright Stars in Globular Clusters}

Analyses of hot UV bright stars in globular clusters uncovered a lack of 
genuine post-AGB stars -- we found only post-early AGB and post-EHB stars. This
may be an explanation for the lack of planetary nebulae in globular clusters
seen by Jacoby et al. (1997). Abundance analyses of post-AGB stars in two
globular clusters suggest that gas and dust separate during the AGB phase.

\subparagraph{Acknowledgements} 

I want to thank M. Catelan, K.S. de Boer, U. Heber, W.B. Landsman, M. Lemke,
R. Napiwotzki, S. Ortolani, and A.V. Sweigart for their collaboration on these
projects. Thanks go also to the staff of the La Silla, ESO, and Calar Alto,
DSAZ, and HST observatories for their support during and after observations. I
gratefully acknowledge support for this work by the DFG (grants Mo~602/1,5,6),
DARA (grant 50~OR~96029-ZA), Alexander von Humboldt-Foundation (Feodor Lynen
fellowship) and Dr. R. Williams as director of the Space Telescope Science
Institute (DDRF grant). 

\subsection*{References}
\litanf

\item Adams S., Seaton M.J., Howarth I.D., Auri\'ere M., Walsh J.R., 1984,
    MNRAS 207, 471
\item Alcaino G., Buonanno R., Caloi V., Castellani V., Corsi C.E., 
    Iannicola G., Liller W., 1987, AJ 94, 917 
\item Arp H.C., 1955, AJ 60, 317
\item Barnard E.E., 1900, ApJ 12, 176
\item Bell R.A., Briley M.M., Norris J.E., 1992, AJ 104, 1127
\item Bergeron, P., Saffer, R.A., Liebert, J., 1992, ApJ 394, 228
\item Bond H.E., 1991, in {\it Evolution of Stars: the Photospheric 
    Abundance Connection}, eds.\ G.~Michaud, A.~Tutukov, IAU Symp. 145 
    (Kluwer: Dordrecht) p.~341
\item Bragaglia, A., C. Cacciari, C., Carretta, E., Fusi Pecci, F., 1999, in 
    {\it The 3$^{rd}$ Conf. on Faint Blue Stars}, eds. A.G.D. Philip, J. 
    Liebert \& R.A. Saffer (Cambridge:CUP), p. 447
\item Brown J.A., Wallerstein G., Cunha K., Smith V.V., 1991, A\&A 249, L13
\item Brown J.A., Wallerstein G., 1993, AJ 106, 133
\item Brown T.M., Ferguson H.C., Davidsen A.F., Dorman B., 1997, ApJ 482, 
    685
\item Buonanno R., Corsi C.E., Fusi Pecci F., 1981, MNRAS, 196, 435 
\item Buonanno R., Buscema G. Corsi C., Iannicola G., Fusi Pecci F., 1983a, 
    A\&AS 51, 83 
\item Buonanno R., Buscema G. Corsi C., Iannicola G., Smriglio F., 1983b, 
    A\&AS 53, 1 
\item Buonanno R., Corsi C.E., Fusi Pecci F., Alcaino G., Liller W., 1984, 
    A\&AS 57, 75 
\item Buonanno R., Caloi V., Castellani V., Corsi C.E., Fusi Pecci F., 
    Gratton R., 1986, A\&AS 66, 79 
\item Buonanno R., Corsi C.E., Buzzoni A., Cacciari C., Ferraro F.R., Fusi 
    Pecci F., 1994, A\&A 290, 69
\item Cacciari C., Fusi Pecci F., Bragaglia A., Buzzoni A., 1995, A\&A 301, 684
\item Caloi V., 1972, A\&A 20, 357
\item Caloi V., 1999, A\&A in press
\item Carretta E., Gratton R.G. 1997, A\&AS 121, 95
\item Catelan M., Borissova J., Sweigart A.V., Spassova N., 1998, ApJ 494, 265
\item Cavallo R.M., Sweigart A.V., Bell R.A., 1998, ApJ 492, 575
\item Code A.D., Welch G.A., 1979, ApJ 228, 95
\item Conlon E.S., 1994, in {\it Hot Stars in the Galactic Halo}, eds. S.
    Adelman, A. Upgren, C.J. Adelman, CUP, p. 309
\item Conlon E.S., Dufton, P.L., Keenan, F.P., 1994, A\&A 290, 897
\item Cool A.M., Piotto G., King I.R., 1996, ApJ 468, 655
\item Crocker D.A., Rood R.T., O'Connell R.W., 1988, ApJ 332, 236
\item d'Cruz N.L., Dorman B., Rood R.T., O'Connell R.W., 1996, ApJ 466, 359
\item de Boer K.S., 1982, A\&AS 50, 247
\item de Boer K.S., 1985, A\&A 142, 321
\item de Boer K.S., 1987, in {\it The 2$^{nd}$ Conference on Faint Blue Stars},
    eds. A.G.D. Philip, D.S. Hayes, J. Liebert, Davis Press, p.~95
\item de Boer K.S., Schmidt J.H.K., Heber U., 1995, A\&A 303, 95 
\item Dickens R.J., Croke B.F.W., Cannon R.D., Bell R.A., 1991, Nature 351, 
    212
\item Dixon W.V., Davidsen A.F., Dorman B., Ferguson H.C., 1996, AJ 111, 1936
\item Dixon W.V., Hurwitz M., 1998, ApJ 500, L29
\item Djorgovski S., 1993, in {\it Structure and Dynamics of Globular Clusters},
    eds. S.G. Djorgovski \& G. Meylan, ASP Conf. Ser. 50, p. 373
\item Dorman, B., Lee Y.-W., VandenBerg D.A., 1991, ApJ 366, 115
\item Dorman, B., Rood, R.T., O'Connell, W.O., 1993, ApJ 419, 596
\item Dorman B., Shah R.Y., O'Connell R.W., Landsman W.B., Rood R.T., et al., 
     1997, ApJ 480, L31
\item Durrell P.R., Harris W.E., 1993, AJ 105, 1420
\item Faulkner J., 1966, ApJ 144, 978
\item Ferraro F.R., Paltrinieri B., Fusi Pecci F., Dorman B., Rood R.T.,
    1998, ApJ 500, 311
\item Fusi Pecci F., Ferraro F.R., Bellazzini M., Djorgovski S., Piotto G., 
    Buonanno R., 1993, AJ 105, 1145
\item Gingold R.A., 1976, ApJ 204, 116
\item Gonzalez G., Wallerstein G., 1994, AJ 108, 1325
\item Gratton R.G., Fusi Pecci F., Carretta E., Clementini G., Corsi C.E.,
    Lattanzi M. 1997, ApJ 491, 749
\item Greenstein J.L., 1939, ApJ 90, 387
\item Greenstein J.L., 1971, in {\it White Dwarfs}, ed. W.J. Luyten,
    IAU Symp. 42, (Reidel), p. 46
\item Greggio L., Renzini A., 1990, ApJ 364, 35
\item Harris H.C., Nemec J.M., Hesser J.E., 1983, PASP 95, 256
\item Heber U., 1983, A\&A 118, 39
\item Heber U., 1987, Mitt. Astron. Ges. 70, 79
\item Heber U., Kudritzki R.P., Caloi V., Castellani V., Danziger J., Gilmozzi
    R., 1986, A\&A 162, 171
\item Heber U., Kudritzki R.P., 1986, A\&A 169, 244
\item Heber U., Dreizler S., Werner, K., 1993, Acta Astron. 43, 337
\item Heber U., Moehler S., Reid I.N., 1997, in {\it HIPPARCOS Venice '97},
    ed. B. Battrick, ESA-SP 402, p. 461 
\item Heber U., Napiwotzki R., 1999, in {\it The 3$^{rd}$ Conf. on Faint
    Blue Stars}, eds. A.G.D. Philip, J. Liebert \& R.A. Saffer 
    (Cambridge:CUP), p. 439 ({\tt astro-ph/9809129})
\item Herwig F., Bl\"ocker T., Sch\"onberber D., El Eid M., 1997, A\&A 324, 
    L81
\item Hoyle F., Schwarzschild M., 1955, ApJS 2, 1
\item Iben I. Jr., 1990, ApJ 353, 215
\item Iben I.Jr., Rood R.T., 1970, ApJ 161, 587
\item Iben I. Jr., Tutukov A.V., 1984, ApJS 54, 335
\item Iben I. Jr., Tutukov A.V., 1993, ApJ 418, 343
\item Jacoby G.H., Morse J. A., Fullton L.K., Kwitter K.B., Henry R.B.C,
    1997, AJ 114, 2611
\item Jordan S., Koester D., A\&AS 65, 367
\item Kraft R.P., 1994, PASP 106, 553 
\item Kraft R.P., Sneden C., Langer G.E., Shetrone M.D., Bolte M., 1995, AJ 
    109, 2586
\item Kraft R.P., Sneden C., Smith G.H., Shetrone M.D., Langer G.E., 
    Pilachowski C.A., 1997, AJ 113, 279
\item Kurucz R.L., 1979, ApJS 40, 1
\item Kurucz R.L., 1992, in {\it The Stellar Populations of Galaxies},
    eds. B. Barbuy \& A. Renzini, IAU Symp. 149 (Kluwer:Dordrecht), 225
\item Landsman W.B., Sweigart A.V., Bohlin R.C., Neff S.G., O'Connell R.W.,
    et al., 1996, ApJ 472, L93
\item Lee Y.-W., Demarque P., Zinn R., 1994, ApJ 423, 248
\item Leone F., Manfr\`e M., 1997, A\&A 320, 257
\item Liebert J., Saffer R.A., Green E.M., 1994, AJ 107, 1408
\item Mathis J.S., Lamers, H.J.G.L.M., 1992 A\&A 259, L39
\item McCausland R.J.H., Conlon E.S., Dufton P.L., Keenan F.P., 1992, ApJ 394, 
    298
\item Moehler S., Heber U., de Boer K.S., 1995, A\&A 294, 65 
\item Moehler S., Heber U., Durrell P., 1997a, A\&A 317, L83
\item Moehler S., Heber U., Rupprecht G., 1997b, A\&A 319, 109
\item Moehler S., Landsman W., Napiwotzki R., 1998a, A\&A 335, 510
\item Moehler S., Heber U., 1998, A\&A 335, 985
\item Moehler S., Heber U., Lemke M., Napiwotzki R., 1998b, A\&A 339, 537
\item Napiwotzki R., 1997 A\&A 322, 256
\item Napiwotzki R., Heber U., K\"oppen, J., 1994, A\&A 292, 239
\item Newell E.B., 1973, ApJS 26, 37
\item O'Connell R.W., Dorman B., Shah R.Y., Rood R.T., Landsman W.B., et al., 
    1997, AJ 114, 1982
\item Paczynski B., 1971, Acta Astron. 21, 1
\item Paltoglou G., Norris J.E., 1989, ApJ 336, 185
\item Parise R.A., Bohlin R.C., Neff S.G., O'Connell R.W., Roberts M.S., 
    et al., 1998, ApJ 501, L67
\item Pease F.G., 1928, PASP 40, 342
\item Peterson R.C., Rood R.T., Crocker D.A., 1995, ApJ 453, 214
\item Pilachowski C.A., Sneden C., Kraft R.P., Langer G.E., 1996, AJ 112, 545
\item Pont F., Mayor M., Turon C., VandenBerg D.A. 1998, A\&A 329, 87
\item Reid I.N. 1997, AJ 114, 161
\item Reid I.N. 1998, AJ 115, 204
\item Renzini A., Voli, M., 1981, A\&A 94, 175
\item Renzini A., Bragaglia A., Ferraro F.R., Gilmozzi R., Ortolani S., 
    et al., 1996, ApJ 465, L23
\item Rich R.M., Sosin C., Djorgovski S.G., Piotto G., King I.R., et al., 
    1997, ApJ 484, L25
\item Richer H.B., Fahlmann G.G., Ibata R.A., Stetson P.B., Bell R.A., 
    et al., 1995, ApJ 451, L17
\item Richer H.B., Fahlmann G.G., Ibata R.A., Pryor C., Bell R.A., 
    et al., 1997, ApJ 484, 741
\item Rood R.T., 1973, ApJ 184, 815
\item Rood R.T., Crocker D.A., 1985, in {\it Horizontal-Branch and 
    UV-Bright Stars''}, ed. A.G.D. Philip (Schenectady: L.Davis Press), p. 99
\item Saffer R.A., Bergeron P., Koester D., Liebert J., 1994, ApJ 432, 351
\item Sandage A.R., Wallerstein G., 1960, ApJ 131, 598
\item Savage B.D., Mathis F.S., 1979, ARAA 17,73
\item Sch\"onberner D., 1983, ApJ 272, 708
\item Schwarzschild M., H\"arm R., 1970, ApJ 160, 341
\item Shapley H., 1915a, Contr. Mt. Wilson 115
\item Shapley H., 1915b, Contr. Mt. Wilson 116
\item Shapley H., 1930, {\it Star Clusters}, The Maple Press Company, York, 
    PA, USA, 
\item Shetrone M.D., 1996, AJ 112, 1517
\item Shetrone M.D., 1997, in {\it Fundamental Stellar Properties: The 
    Interaction between Observation and Theory}, IAU Symp. 189 (poster 
    proceedings) (Kluwer: Dordrecht), p. 158
\item Smith V.V., Cunha K., Lambert D.L., 1995, AJ 110, 2827
\item Smith G.H., Shetrone M.D., Bell R.A., Churchill C.W., Briley M.M., 1996, 
    AJ 112, 1511
\item Sneden C., Kraft R.P., Prosser C.F., Langer G.E., 1992, AJ 104, 2121
\item Sneden C., Kraft R.P., Shetrone M.D., Smith G.H., Langer G.E., 
    Prosser C.F., 1997, AJ 114, 1964
\item Sosin C., Piotto G., Djorgovski S.G., King I.R., Rich R.M., Dorman B.,
    Liebert J., Renzini A., 1997, in {\it  Advances in Stellar Evolution},
    eds. R.T. Rood \& A. Renzini, CUP, p. 92
\item Stecher T., Cornett R.H., Greason M.R., Landsman W.B., Hill J.K.,
    et al., 1997, PASP 109, 584
\item Stoeckley R., Grennstein J.L., 1968, ApJ 154, 909
\item Strom S.E., Strom K.M., 1970, ApJ 159, 195
\item Strom S.E., Strom K.M., Rood R.T., Iben I.Jr., 1970, A\&A 8, 243
\item Sweigart A.V., 1987, ApJS 65, 95
\item Sweigart A.V., 1994, in {\it Hot Stars in the Galactic Halo}, eds. S.
    Adelman, A. Upgren, C.J. Adelman, CUP, p. 17
\item Sweigart A.V., 1997, ApJ 474, L23
\item Sweigart A.V. 1998, to appear in {\it New Views on the Magellanic 
    Clouds}, eds. Y.-H. Chu, J. Hesser \& N. Suntzeff, IAU Symp. 190 (ASPC)
\item Sweigart A.V. 1999, in {\it The 3$^{rd}$ Conf. on Faint
    Blue Stars}, ed. A.G.D. Philip, J. Liebert \& R.A. Saffer 
    (Cambridge:CUP), 3 ({\tt astro-ph/9708164})
\item Sweigart A.V., Mengel J.G., Demarque P., 1974, A\&A 30, 13
\item Sweigart A.V., Gross P.G., 1974, ApJ, 190, 101
\item Sweigart A.V., Gross P.G., 1976, ApJS 32, 367
\item Sweigart A.V., Mengel J.G., 1979, ApJ 229, 624
\item Sweigart A.V., Catelan M., 1998, ApJ 501, L63
\item ten Bruggencate P., 1927, {\it Sternhaufen}, Julius Springer Vlg., 
    Berlin
\item Traving G., 1962, ApJ 135, 439
\item Tuchman Y., 1985, ApJ 288, 248
\item Vassiliadis E., Wood P.R., 1993, ApJ 413, 641
\item Whitney J.H., Rood R.T., O'Connell R.W., D'Cruz N.L., Dorman B., 
    et al., 1998, ApJ 495, 284
\item Yi S., Demarque P., Kim Y.-C., 1997, ApJ 482, 677
\item Yi S., Demarque P., Oemler A. Jr., 1998, ApJ 492, 480
\item Zinn R., 1974, ApJ, 193, 593
\item Zinn R.J., Newell E.B., Gibson J.B., 1972, A\&A 18, 390
\item Zinn R., West M.J. 1984, ApJS 55, 45
\litend
\end{document}